\documentclass[preprint]{aastex}
\begin{document}

\title{Age Determination for 346 Nearby Stars in the Herschel DEBRIS Survey}

\author{Laura Vican\altaffilmark{1}}

\altaffiltext{1}{Department of Physics and Astronomy, University of California, Los Angeles, CA 90095, USA (lvican@astro.ucla.edu)}

\begin{abstract}
DEBRIS is a flux-limited survey of nearby stars (spectral types A-M) for evidence of debris disks with the {\it Herschel Space Observatory}. One goal of the survey is to determine disk incidence as a function of various stellar parameters. Understanding debris disk evolution depends on knowledge of the precise age of stars around which these debris disks are found. However, finding ages for field stars is notoriously difficult. Furthermore, in an unbiased sample like DEBRIS, one is working with stars across many spectral types. This requires a multi-method approach to age determination. In this paper, we outline several methods of age determination broken down by spectral type, including some strengths and limitations of each method. In total, we were able to calculate ages for 263 of 274 F, G, and K-type stars, and all 83 A-type stars in the DEBRIS sample. 

\end{abstract}


\clearpage

\section{Introduction}
Debris disks represent an important aspect of the evolution of extra-solar planetary systems. By studying the formation and evolution of debris disks, we can learn a great deal about the formation and evolution of our solar system (Wyatt 2008, Zuckerman 2001). Debris disks consist of small grains of dust that have been ground down by eons of collisions between rocky bodies in orbit around the central star. It is thought that our own solar system went through an especially prominent debris disk phase during the Late Heavy Bombardment (Booth et al. 2009). The dust grains produced by violent collisions of rocky bodies can be observed via thermal emission from infrared to centimeter wavelengths. For a more detailed analysis of circumstellar disks, see Heng $\&$ Tremaine (2010). 

The DEBRIS (Disk Emission via a Bias-free Reconnaissance in the Infrared/Submillimeter) project aims to characterize the types of stars which host circumstellar disks using the {\it Herschel Space Observatory}, hereafter {\it Herschel}. The survey includes a roughly equal number of stars from each spectral type A-M, with 446 stars in total (Phillips et al. 2010, Matthews et al. 2010). DEBRIS is an unbiased survey, meaning that it did not specifically target stars with observed IR excesses. We want to know how debris disk incidence trends with various stellar parameters, and how those parameters evolve with time. Our motivation for determining the ages of stars is to better characterize the evolution of these disks.

A concerted effort has been made in the past five years to derive age determination equations based on a variety of parameters. In Section 2, we discuss the age determination methods available for solar type (F, G, and K-type) stars. In Section 2.1, we explore the possibility of gyrochronology, the calculation of stellar age based on the spin-down of a star over its lifetime. In Section 2.2, we discuss chromospheric emission, the use of the evolution of a star's magnetic field as an age tracer. In Section 2.3, we discuss the derivation of stellar age using X-ray emission. In 2.4, we briefly discuss the use of lithium depletion as an age tracer. In Section 2.5, we compare four age determination methods (gyrochronology, chromospheric, X-ray, vsin$i$) against each other. In 2.6, we discuss which ages represent the closest estimate of the ``true" age.  In Section 3.1, we discuss the use of isochrone dating in A-type stars. Section 3.2 describes other possible methods for A star age determination. M-type stars are not treated in this paper, as their evolutionary properties are currently not well understood, and few of them have been found to host debris disks. 

\section{\large F, G, and K-type stars}
\subsection{Method 1: Gyrochronology}
Developed by Barnes (2003), this method has been used by many researchers in recent years to determine relatively precise ages for field stars. The basic idea is that magnetic winds carry angular momentum away from the star, causing the outer convective envelope to spin slower than the radiative zone beneath. This creates a shear, which slows down the overall rotation of the star with time. This period change is predictable, and can be calibrated using the known age of the Sun. Barnes (2007) suggested the following relation between age (t), rotation period (P$_{rot}$), and B-V color:

\begin{equation}
P_{rot}(B-V,t)=a[(B-V)_{0}-c]^{b}t^{n}
\end{equation}

Mamajek $\&$ Hillenbrand (2008, hereafter MH08) calibrated this relation using four clusters with known ages ($\alpha$ Per - 85 Myr, Pleiades - 130 Myr, M34 - 200 Myr, and Hyades - 625 Myr). They found that a=0.407, b=0.325, c=0.495, and n=0.566 for t in Myrs and P$_{rot}$ in days. This relation could only be calibrated for stars with  0.495$<$(B-V)$_{0}$$<$1.4 and was not fit to any clusters older than the Hyades due to lack of rotational data in the literature. Therefore, their low quoted error of $\sim$15 $\%$ is only valid for ages less than 625 Myr. While no cluster data were available beyond this age limit, MH08 showed that the relation recovered similar ages for both components of binary systems out to 3 Gyr (with 20-25 $\%$ error). The error in age for stars in our sample older than $\sim$1 Gyr is taken to be $\sim$20$\%$.

Several studies use short term (on the order of weeks) variations in the strength of Ca II H \& K emission cores to measure rotation period (Donahue et al. 2006). Ca II emission variations trace the rotation of areas of increased magnetic flux (starspots) across the surface of the star. Thus, these fluxuations likely represent a true rotational period. These variations differ from the longer term (on the order of Myrs) magnetic variations discussed in Section 2.2.

Gyrochronology using true rotational periods was possible for 35  F, G, and K type DEBRIS stars.  We gathered rotation period data from various sources (Baliunas et al. 1996 - 5 stars, Baliunas et al. 1983 - 2 stars, Donahue et al. 1996 - 16 stars, Wright et al., 2011 - 12 stars) and used B-V color data from the Hipparcos catalog (Perryman et al. 1997).  Table 1 contains our age results and the relevant parameters. Figure 1 plots B-V vs. P$_{rot}$ for these data with so-called $gyrochrones$ (using Equation 1 with coefficient values from MH08) plotted for 0.3, 2, 5, and 7 Gyr.

Gyrochronology is less efficient for dating younger stars, since initial conditions (such as initial rotational period) become less important over time. There is also a recognized rotation period saturation limit above a mass-dependent maximum rotational period (Barnes 2010). Beyond this limit (t$_{sat}$), rotation and age cease to be related by the gyrochronology equation of MH08. For a solar twin (B-V$\sim$0.65), the rotation saturation occurs $\sim$7 Gyr. 

\subsubsection{Using vsin$i$ as a Proxy for Measured Rotation Periods}

To use the gyrochronology equation to calculate age, one must know the rotation period of a star. However, rotation is very difficult to measure directly, especially for slowly rotating stars. Most ``rotation rates" in the literature are actually vsin$i$ measurements taken from spectroscopic observations (see Rutten et al. 1987, Valenti et al. 2005). One might estimate a rotation period from a vsin$i$ measurement by assuming that the average inclination ($i$) over a sample should be $\pi$/2. We tested the validity of using vsin$i$ to estimate ages by calculating ages of stars using rotational period (P$_{rot}$) and vsin$i$. We found that vsin$i$-based age calculations are not well-correlated with P$_{rot}$-based age calculations (see Figure 2). On average, the ages derived using vsin$i$ differed from the ages derived from P$_{rot}$ by $\sim$98$\%$ (see Figure 2). We therefore prefer gyrochronology ages based on P$_{rot}$ over those based on vsin$i$. We were able to find vsin$i$ ages for 113 stars. Those data are presented in the right-hand column of Table 2. 

Direct comparison of the measured rotational period to the rotational period predicted by vsin$i$ showed reasonably good agreement between the two at short rotational period (<15 days); the two periods (measured and calculated) disagree on average by $\sim$45$\%$. The agreement at short rotational period is likely a result of the fact that both vsin$i$ and rotational period are easier to measure for short rotational period (i.e. high velocity). That is, stars with short rotation periods are easy to monitor photometrically. In addition, stars with high rotational velocities are easy to measure spectroscopically. 

\subsection{Method 2: Chromospheric Activity}

This method was first introduced by Baliunas et al. (1983) and used by MH08 to calculate the age of field stars. The physics behind chromospheric emission is an ongoing topic of research in solar and extra-solar astronomy.  Angular momentum transport through convection is more efficient than through radiation. Thus, for stars with an outer convective envelope, the outer layer spins down faster than the inner layer, creating a shear at the interface between the layers. This shear induces a magnetic field that is then perturbed and twisted by convective motions. The magnetic field lines carry this stress up into the chromosphere where magnetic heating causes emission lines to form in the cores of Ca II H and K absorption lines. Since magnetic activity decreases with stellar age, this emission will also decline as a function of stellar age. This method has been used successfully to recover the age of stars in clusters with known ages (MH08, Wright et al. 2004). Chromospheric heating is characterized by the parameter $R^\prime_{HK}$; the larger is $R^\prime_{HK}$ the more active - and therefore younger - is a given star.

Many researchers have taken optical spectra, measured the strength of the Ca II H $\&$ K lines and published $R^\prime_{HK}$ values (Soderblom et al. 1985, Henry et al. 1996, Gray et al. 2003, Maldonado et al. 2010, Wright et al. 2004, and Gray et al. 2006). We used these $R^\prime_{HK}$ values to calculate chromospheric ages for stars in the DEBRIS sample with the following formula from MH08 (for t in years):

\begin{equation}
log(t)=-38.05-17.91\ log(R'_{HK})-1.67\ log(R'_{HK})^{2}
\end{equation}

MH08 derived this empirical relation by fitting a curve to stars in clusters with known ages out to 4 Gyr. The applicability of this equation is limited by the sample of clusters used for the fit, and is therefore only accurate between 

\begin{displaymath}
-5.1<log(R'_{HK})<-4.0
\end{displaymath} 

This corresponds to an age range of $\sim$0.05 Gyr to 6 Gyr for a solar twin (B-V$\sim$0.65). 

Wright et al. 2004 (hereafter W04) provide an equation for calculating ages from chromospheric activity which originally came from Donahue (1993). For stars in common, we derive similar ages to MH08 (see Figure 3), using the more recent equation from MH08. While there is a close correlation between the ages, it would seem that for stars older than 2 Gyr, the MH08 relation (Equation 2) produces older ages than W04, whereas for young stars ($<$2 Gyr), the opposite may be the case. This discrepancy was discussed at length in Song et al. (2004) (hereafter S04). S04 derived ages for stars less than a few hundred Myrs old using lithium abundance (discused in Section 2.4), X-ray activity (discussed in Section 2.3), and Galactic UVW space motions. Comparing those ages to ages derived using the relation from W04, S04 found that chromospheric ages from W04 were systematically older than ages derived using other methods. Discrepancies such as this are a strong motivator for revised activity-age relations, and a reason that we choose to use the more modern relation from MH08 to calculate ages for our sample; in comparison with the W04 chromospheric ages, young star chromospheric ages from MH08 are in better agreement with the S04 ages.

A total of 255 chromospheric ages were determined for DEBRIS stars with 35 stars overlapping with our calculated gyrochronology ages. A plot of gyrochronology age vs. chromospheric age can be found in Figure 4 and shows moderately good agreement between gyrochronology ages and chromospheric ages. MH08 quotes a typical error for chromospheric ages of 60$\%$.

There is an intrinsic limitation in using magnetic activity as an age tracer. We know that our Sun undergoes an 11 year activity cycle during which its log($R^\prime_{HK}$) changes from -4.83 during minimum to -4.96 during maximum (MH08). It is thought that solar-type stars undergo similar activity cycles. Using snapshot spectra of field stars, it is not possible to tell whether the star is in an activity minumum or maximum. In the case of the Sun, its calculated age at minimum and maximum is 3.64 Gyr and 5.85 Gyr respectively. This still gives a reasonable estimate of the solar age (4.57 Gyr). Thus, while multiple epoch surveys are preferable, we can still use single epoch surveys to estimate stellar age.

The only way to determine the $average$ magnetic activity over a stellar cycle (which is the parameter that actually decreases with age) is to conduct a long-term observation campaign. Such observations were made for 1296 stars at Mount Wilson observatory over a 17 year period from 1966-1983 (Duncan et al. 1991). In this paper, mean ``S-values" are listed for each star. These S values represent emission flux densities in Ca II H $\&$ K lines, and can be converted to $R^\prime_{HK}$.

When the average activity level of a star is used in the calculation of $R^\prime_{HK}$, the error due to stellar magnetic variation on short timescales should be significantly reduced. Two of our sources of $R^\prime_{HK}$ made efforts to take this variability into account in their measurements. W04 used median activity levels from 6 years of observation at Keck Observatory and 17 years of observation at Lick Observatory to calibrate their published $R^\prime_{HK}$ values. Duncan et al. (1991) contains Mount Wilson data as described above. When possible, we took our $R^\prime_{HK}$ values from one of these two sources. Otherwise, we used $R^\prime_{HK}$ values from single-epoch observing papers such as Gray et al. (2003 $\&$ 2006), Maldonado et al. (2010), and Henry et al. (1996). For stars in Table 2 observed with multi-epoch surveys, the chromospheric age should be trusted before the X-ray ages, since the X-ray age relation was derived from the chromospheric age relation (X-ray ages are described in detail in Section 2.3). 

With the advent of long-term monitoring of solar-type stars (mostly in search of exoplanet transits), it will soon be possible to quantify this magnetic variation for a larger fraction of solar-type stars.

\subsection{Method 3: X-Ray Emission}
X-rays trace magnetic heating of the stellar corona. Although it is not well known exactly how the corona is heated, MH08 suggests that it is closely related to the strength of the magnetic field, and also to chromospheric heating.  

X-ray count rates and hardness ratios (HR1) for the 0.1-2.4 keV band are readily available from the ROSAT survey (Voges et al. 1999). We first calculated the X-ray luminosity for 100 DEBRIS stars by the following equation from MH08:

\begin{equation}
 L_x = 4\pi D^2 C_x f_x
\end{equation}

\noindent where D is distance in cm, f$_x$ is the ROSAT count rate (counts s$^{-1}$), and C$_x$ is a conversion factor defined by the following equation from MH08:

\begin{equation}
C_x=(8.31+5.3HR1) \times 10^{-12}
\end{equation}

Using stellar radii and T$_{eff}$ values from several sources (Allende Prieto et al. 1999, Valenti et al. 2005, Takeda et al. 2007), we were able to calculate L$_{bol}$ from L$_{bol}$=4$\pi$R$^{2}$$\sigma$T$_{eff}^{4}$. MH08 fit a relation between log($R^\prime_{HK}$) and log(L$_{x}$/L$_{bol}$) to derive an age relation for X-ray activity:

\begin{equation}
log(t)=1.20-2.307\ log(R_{x}) - 0.1512\ log(R_{x})^{2}
\end{equation}

\noindent where t is in years and R$_{x}$=L$_{x}$/L$_{bol}$. In MH08, this relation was never fit to cluster data. Rather, R$_{x}$ was used to predict $R^\prime_{HK}$ by:

\begin{equation}
logR'_{HK}=-4.54+0.289\ (log(R_x)+4.92)
\end{equation}

This relation was combined with Equation 2 and used to derive Equation 5. Therefore we expect scatter in the R$_x$ ages (when compared to gyrochronology ages) both from uncertainty of the actual X-ray count measurements, and from the scatter in the R$_x$ - $R^\prime_{HK}$ relation (which MH08 estimates at 55$\%$). X-ray ages and relevant parameters can be found in Table 2. We compared ages determined using X-ray flux to ages determined by gyrochronology. The results are shown in Figure 5. A clear correlation is apparent, with an average scatter of $\sim$84$\%$. Since gyrochronology produces ages with the smallest error, we can conclude that X-rays are legitimate tracers of stellar age. 

While X-rays are useful for determining the ages of young, active stars, there is a saturation limit beyond which age and magnetic activity cease to be correlated by Equation 5. This ``saturation" limit defines a minimum age (and therefore maximum magnetic activity level) which can be reliably determined by the methods in this paper. The actual cutoff for this limit varies between studies. According to MH08, the relation between R$_{x}$ and $R^\prime_{HK}$ ceases to correlate above logR$_{x}$=-4. However Zuckerman $\&$ Song (2004) find moderately tight isochrones for young nearby stars and the Pleiades cluster in logR$_{x}$-color space above logR$_{x}$=-4. They do place a lower limit on the age which can reliably determined using X-rays ($\sim$100 Myr). 

\subsection{Lithium Depletion}
This method was used by Baumann et al. (2010) to determine the age of planet-hosting stars. Lithium is depleted from the photosphere by way of convective mixing into interior regimes where the Li is burned. This process is dependent on stellar rotation; The faster a star rotates, the slower is this process, since there is less differential rotation and thus less mixing. There exist isochrones in lithium abundance - rotation velocity space which can be used to determine the age of a star in isolation (e.g., Chen et al. 2001, Zuckerman $\&$ Song 2004). Since rotation period is also necessary for age-determination using lithium depletion, we prefer to use gyrochronology - which has a direct equation for determining age - rather than the less precise method of matching isochrones.

In addition, the accuracy of the derived age using lithium depletion varies depending on spectral type. It is most accurate for late-K to early M-type stars. Even so, the isochrones are not well-defined in EW(Li) - age space (see Figure 3 of Zuckerman $\&$ Song 2004). Ages determined using litium abundances are certainly not as accurate as our other methods. We therefore choose not to use lithium depletion to date stars in the DEBRIS survey. 

\subsection{Comparing Four Age Determination Methods}
We began our analysis by assuming that gyrochronology provides the most accurate estimation of stellar age. We then compared the other three methods (chromospheric activity, X-ray emission, and vsin$i$ measurements) to gyrochronology in a quantitative way. Two stars were left out of the analysis; HIP 19849 and HIP 25647. HIP 19849 is an extremely slow rotator - thus, its gyrochronology age may not be accurate. Also, HIP 25647 has a very low gyrochronology age (2 Myr). Its gyrochronology age is unreliable as well.

\bf Chromospheric Ages vs. Gyrochronology Ages\rm: There are 35 stars in the DEBRIS sample for which we have both chromospheric ages and gyrochronology ages. We disregard the two ourliers mentioned above and restrict our analysis to the remaining 33 stars. We define a ``discrepancy" factor, f$_{c}$, by:

\begin{displaymath}
f_{c}=\frac{t_{chromo}}{t_{gyro}}
\end{displaymath}

For the sample of 34 stars, the standard deviation ($\sigma_{c}$) of f$_{c}$ was 0.872 around a median discrepancy of f$_{c,m}$=1.15 (and a mean of 1.32). This includes ages derived using $R^\prime_{HK}$ values from both single and multiple epoch observing campaigns.\footnotemark{} 

\footnotetext{This analysis should only be used as a comparative tool, and cannot be used to reliably determine the errors associated with these age determination methods (since we do not know the true age of the star). When this method was used to predict ages of stars in clusters (MH08) it yielded a 60$\%$ error. We take this to be the typical error associated with chromospheric age dating.} 

\bf X-Ray Ages vs. Gyrochronology Ages\rm: There are 17 stars in the DEBRIS sample for which we have both X-ray ages and gyrochronology ages. Again, we disregard HIP 19849 and HIP 25647.  For the remaining 16 stars, we again defined a discrepancy factor, f$_{x}$, by:

\begin{displaymath}
f_{x}=\frac{t_{xray}}{t_{gyro}}
\end{displaymath}

We found that $\sigma_{x}$=0.518 and f$_{x,m}$=0.89 (the mean of the discrepancies was found to be 1.32). This seems to suggest that X-ray ages are more precise than chromospheric ages. 

\bf Chromospheric Ages vs. X-Ray Ages\rm: We compared ages calculated using log(R$_{x}$) to ages calculated using log$R^\prime_{HK}$ to produce Figure 6.  Chromospheric ages are, on average, higher than X-ray ages. Otherwise, there is little to no correlation between the two age determination methods. 

\bf vsin$i$ Ages vs. P$_{rot}$ Ages\rm: We also compared ages calculated using P$_{rot}$ to ages calculated using vsin$i$. We define a discrepancy factor f$_{v}$ by:

\begin{displaymath}
f_{v}=\frac{t_{vsini}}{t_{gyro}}
\end{displaymath}

We found that $\sigma_{v}$=1.6 around a median f$_{v,m}$ of 0.94. 

Since the same gyrochronology equation is used to calculate age in both cases, what we really wanted to compare was the true P$_{rot}$ of a star to the P$_{rot}$ calculated from vsin$i$. For this purpose, we define a discrepancy factor f$_{p}$ by:

\begin{displaymath}
f_{p}=\frac{P_{vsini}}{P_{rot}}
\end{displaymath}

We find that $\sigma_{p}$=0.66 around a median f$_{p,m}$ of 0.97. The scatter in the relationship between ages calculated using vsin$i$ and ages calculated using P$_{rot}$ is largely due to the scatter in the relationship between the periods measured using vsin$i$ and the measured rotational period. From this analysis, we conclude that while vsin$i$ may be used to predict a rotational period with an error of $\sim$60$\%$, the scatter that is introduced into the gyrochronology equation from this prediction is quite high. We suggest that chromospheric and X-ray ages be taken to be more precise than vsin$i$ ages. 

\subsection{Which Age Should be Used?}
In general, the gyrochronology age should be trusted above any other available age (except in the two cases mentioned above - HIP 19849 and HIP 25647). When the gyrochronology age is unavailable or unreliable, one needs to be able to make a choice between the remaining two methods of age determination. We first define a discrepancy factor f$_{cx}$:

\begin{displaymath}
f_{cx}=\frac{t_{chromo}}{t_{xray}}
\end{displaymath}

There are 62 stars in Table 2 with f$_{cx}$ $<$2. Of the remaining 30 stars, 13 stars have $R^\prime_{HK}$ values taken from multi-epoch surveys. Although our analysis in Section 2.4 suggested that X-ray ages agree better with gyrochronology ages, we know that X-ray ages depend on snapshot observations of X-ray emission. This means that ages derived from X-ray observations will be affected by the short-term variability of the stellar magnetic cycle. Thus, for these 13 stars, we believe the chromospheric ages to be the most accurate. 

Finally, we were left with 17 stars which were derived using $R^\prime_{HK}$ values taken from single-epoch surveys. For these stars, we were able to use other age determination methods such as galactic space motion (Zuckerman $\&$ Song 2004), gyrochronology ages, or isochrone ages from the literature to make an educated choice between the X-ray age and the chromospheric age. 

\bf HIP 13402\rm: This star has a gyrochronology age of 0.256 Gyr. This agrees best with its X-ray age 0.14 Gyr. 

\bf HIP 17420, 86036, 102485\rm: These stars have isochrone ages in the literature which agree best with the derived chromospheric ages. 

\bf HIP 80686, 61174\rm: These stars have isochrone ages in the literature which agree best with the derived X-ray ages.

\bf HIP 104440, 2762, 14879, 85295, 67153\rm: For these stars, we recommend taking the chromospheric age as the true age, because their UVWs (Anderson $\&$ Francis 2011) are most compatible with the chromospheric age.

\bf HIP 73695, 89805, 114948\rm: For these stars, we recommend taking the X-ray age as the true age, because their UVWs (Anderson $\&$ Francis 2011) are most compatible with the X-ray age.

\bf HIP 98470, 67422, 5896\rm:  For these stars, we are unable to choose between their X-ray ages and chromospheric ages. Further long-term observations are needed to constrain their ages.

\subsection{FGK Type Star Summary}

(1) Gyrochronology produces the lowest errors of any of our age determination methods (typical errors of 15-20$\%$). It is most precise for intermediate age stars (100 Myr $<$ t $<$ t$_{sat}$ where t$_{sat}$ is B-V dependent). However, rotation periods are difficult to measure and vsin$i$ values are an inadequate proxy.

(2) Ca II chromospheric emission is a legitimate tracer for stellar age;  $R^\prime_{HK}$ is an easily measured parameter, and is readily available in the literature for many field stars. Chromospheric emission is most precise in the same age regime as gyrochronology (described above). MH08 quotes a typical error of 60$\%$, although our analysis suggests an error of 87$\%$.

(3) X-ray flux is a valid tracer of stellar age. Typical errors in the age are estimated at $\sim$70$\%$ (MH08).

(4) Lithium depletion has been shown to correlate with stellar age in past studies (Chen et al. 2001); however it is most useful for young stars ($<$100 Myr) and requires knowledge of the stellar rotation period (Zuckerman $\&$ Song 2004).

\section{\large A Type Stars}
\subsection{Method: Isochrones in Log(g) - log(T$_{eff}$) Space}
 Since A stars evolve quickly, their ages can be reliably determined (errors of about 100-300 Myr) from their position on an HR diagram. We began with three separate sets of isochrones in log(g) - log(T$_{eff}$) space. The first is from Li $\&$ Han (2008), the second is the Padova set of isochrones (Bertelli et al. 2009) and the third is the YREC set of isochrones (Pinsonneault et al. 2004). Values of log(g) come from a variety of sources (Gray et al. 2003, Gray et al. 2006, Lafrasse et al. 2010, Soubiran et al. 2010, Allende Prieto et al. 1999, King et al. 2003, Gerbaldi et al. 1999, and Song et al. 2001). Values of T$_{eff}$ came from Phillips et al. (2010). We estimated the age of these stars using all three sets of isochrones, then compared those resulting ages to each other and to ages published in the literature. We concluded that the YREC isochrones provide a better match to literature ages. Not all of the A stars in the DEBRIS sample fall in the area of log(g)-T$_{eff}$ space that is covered by the YREC isochrones. In some cases, we were able to use Li $\&$ Han (2008) isochrone ages. We filled in any missing ages using ages from Rieke et al. (2005), which uses Y2 isochrones from Yi et al. (2003). 

In total, we were able to determine ages for all 83 A stars in the DEBRIS survey - 65 from YREC isochrones, 15 from Li $\&$ Han (2008) isochrones, 2 from Rieke et al. (2005), and 1 from cluster membership. HIP 88726 was found to be a member of the Beta Pictoris moving group, and was assigned an age of 0.012 Gyr according to Zuckerman $\&$ Song (2004).  Figure 7 shows a histogram of our derived A star ages (83 stars). 

\subsection{Other Methods for A Stars}
Gyrochronology will not work as well for A-type stars as it does for F, G, and K-type stars. Since A stars evolve quickly, they do not spend as much time on the main sequence, and thus their rotation does not brake in the same way as do solar-types.

Furthermore, according to MH08, the correlation between magnetic activity and age breaks down for (B-V)$_0$ $<$ 0.5. Below this limit (which includes A stars) convective envelopes are thin or nonexistent and therefore magnetic field strength caused by rotational shear between the convective and radiative layers diminishes. We therefore prefer isochrone dating to measure the ages of A stars.

\section{\large Conclusions and Future Work}

In total, we were able to reliably determine ages for $\sim$96$\%$ of A-K type stars in the Herschel DEBRIS project. Our motivation for this work was to provide the ages of DEBRIS target stars as a diagnostic for debris disk evolution. Forthcoming papers (i.e. Thureau et al. 2012, in prep, Sibthorpe et al. 2012, in prep) will elaborate on specific disk characteristics as a function of stellar age.

While DEBRIS was an unbiased survey, we hope to expand our age determination project to include any stars observed by Herschel with observed infrared excesses. 

Partial support for this work, part of the NASA Herschel Science Center Key Program Data Analysis Program, was provided by NASA through a contract (No. 1353184, PI: H. M. Butner) issued by the Jet Propulsion Laboratory, California Institute of Technology under contract with NASA. The program, DEBRIS, or Disc Emission via a Bias-free Reconnaissance in the Infrared/Submillimetre, is a Herschel Key Program (P. I. Matthews).  Herschel is the 4th cornerstone mission of the European Space Agency (ESA) science program.

We thank Ben Zuckerman for his valuable insight, David Rodriguez and Erik Mamajek for their helpful input, and Greg Henry for allowing us access to his rotation period data (Wright et al., in press) prior to publication.

\clearpage

\begin{deluxetable}{cccccc}
\tablewidth{0pt}
\tablecaption{DEBRIS Stars with Gyrochronology Ages}
\tablehead{
\colhead{HIP}&
\colhead{Age}&
\colhead{P$_{rot}$}&
\colhead{P$_{rot}$}&
\colhead{B-V}\\
\colhead{}&
\colhead{(Gyr)}&
\colhead{(days)}&
\colhead{source}&
\colhead{}}
\startdata
171&6.07&33&B96&0.69\\
1803&0.52&7.78&D96&0.659\\
3093&8.33&48&D96&0.85\\
3765&5.31&38.5&W11&0.89\\
7981&4.93&35.2&D96&0.836\\
8768&0.67&15.8&W11&1.424\\
12114&7.53&48&D96&0.918\\
12444&1.3&7.4&W11&0.524\\
13402&0.26&6.76&D96&0.862\\
15457&0.68&9.4&B83&0.681\\
19849&7.22&43&B96&0.82\\
22263&0.55&7.6&B83&0.632\\
25647&0&0.51&W11&0.83\\
42438&0.27&4.9&W11&0.618\\
43726&1.65&15&B96&0.661\\
44897&1.08&9.67&D96&0.585\\
47080&1.81&18.6&D96&0.77\\
55454&0.41&11.6&W11&1.34\\
56242&2.31&14&W11&0.57\\
56997&1.66&16.68&D96&0.723\\
57939&4.61&31&B96&0.754\\
64394&1.83&12.35&D96&0.572\\
64792&0.16&3.33&D96&0.585\\
64797&1.38&18.47&D96&0.926\\
64924&4.58&29&W11&0.709\\
67275&0.47&3.2&W11&0.508\\
72659&0.3&6.31&D96&0.72\\
77257&5.49&25.8&D96&0.604\\
77408&1.03&14.05&W11&0.801\\
88972&6.43&42.4&D96&0.876\\
94346&0.2&5.49&W11&0.804\\
98698&2.44&28.95&D96&1.128\\
107350&0.32&4.86&D96&0.587\\
113283&0.38&9.87&W11&1.094\\
113357&8.01&37&B96&0.666
\enddata
\tablecomments{Col (5): B83 = Baliunas et al. (1983), B96 = Baliunas et al. (1996), D96 = Donahue et al. (1996), W11 = Wright et al. (2011)}
\end{deluxetable}

\begin{deluxetable}{ccccccccc}
\tablewidth{0pt}
\tablecaption{DEBRIS Stars with Activity Ages}
\tablehead{
\colhead{HIP}&
\colhead{B-V}&
\colhead{log($R^\prime_{HK}$)}&
\colhead{$R^\prime_{HK}$}&
\colhead{log(R$_{x}$)}&
\colhead{vsin$i$}&
\colhead{Age$_{chromo}$}&
\colhead{Age$_{X-ray}$}&
\colhead{Age$_{vsini}$}\\
\colhead{}&
\colhead{}&
\colhead{}&
\colhead{source}&
\colhead{}&
\colhead{km/s}&
\colhead{Gyr}&
\colhead{Gyr}&
\colhead{Gyr}}
\startdata
169&1.39&-4.724&G03&&&2.24&&\\
171&0.69&-4.94&D91&&4.07&5.49$^{b}$&&\\
473&1.41&-4.702&G06&&&2.01&&\\
544&0.752&-4.32&M10&-4.35&4.1&0.16&0.24&0.66\\
910&0.487&-4.917&G03&&4.88&5.07&&\\
950&0.459&-4.715&G06&&&2.15&&\\
1292&0.749&-4.44&H96&-4.39&&0.4&0.26&\\
1599&0.576&-4.85&H96&&3&3.95&&3.38\\
1803&0.659&-4.44&W04&-4.58&7&0.4$^{b}$&0.4&0.42\\
2762&0.567&-4.615&S09&-4.45&&1.25&0.3$^{a}$&\\
2941&0.715&-4.94&W04&&5.54&5.49&&0.48\\
3093&0.85&-5.02&W04&&1.1&6.98$^{b}$&&6.15\\
3583&0.635&-4.55&H96&&&0.84&&\\
3765&0.89&-5.25&H96&&2&10.58&&1.48$^{b}$\\
3810&0.501&-5.011&G06&&&6.81&&\\
3850&0.769&-4.72&W04&&1.1&2.2&&5.84\\
3909&0.514&-4.612&G03&-5.83&&1.23&3.26&\\
4022&1.29&-4.628&G06&&&1.34&&\\
4148&0.936&-4.83&M10&&1.8&3.64&&1.60\\
5799&0.448&-4.94&D91&-5.41&13.04&5.49&1.81&\\
5862&0.571&-4.95&H96&&4.3&5.67&&2.40\\
5896&0.48&-5.037&G06&-5.15&&7.29&1.18$^{a}$&\\
7235&0.766&-4.6&W04&&4.88&1.14&&1.29\\
7513&0.536&-5.04&W04&-6&9.6&7.35&3.98&1.30\\
7918&0.618&-5&D91&&4.07&6.6&&1.80\\
7978&0.551&-4.997&G06&&5.6&6.54&&1.50\\
7981&0.836&-4.95&W04&&1.7&5.67$^{b}$&&2.53\\
8497&0.333&-4.626&G06&-5.2&&1.33&1.28&\\
8768&1.424&-4.59&M10&&3.21&1.08$^{b}$&&\\
10138&0.812&-4.68&M10&-5.62&2.4&1.79&2.46&1.41\\
10644&0.607&-4.71&W04&-5.1&3.93&2.09&1.08&1.65\\
10798&0.724&-4.85&W04&&3.26&3.95&&\\
12114&0.918&-5.22&D91&&2.9&10.25$^{b}$&&0.73\\
12444&0.524&-4.69&D91&-5.31&&1.89$^{b}$&1.54&\\
12530&0.51&-4.383&S09&-4.56&&0.26&0.38&\\
12653&0.561&-4.65&H96&-5.01&6.5&1.52&0.91&1.14\\
12843&0.481&-4.458&G06&-4.79&28.02&0.46&0.6&\\
13402&0.862&-4.504&M10&-4.12&4.9&0.62$^{b}$&0.14$^{a}$&0.33\\
14445&1.358&-3.88&G06&&&0.002&&\\
14879&0.543&-4.901&G06&-4.58&4.41&4.79&0.39$^{a}$&6.85\\
14954&0.575&-4.88&W04&&8.6&4.44&&1.21\\
15371&0.6&-4.79&H96&&2.6&3.06&&3.24\\
15457&0.681&-4.47&D91&-4.64&5.2&0.5$^{b}$&0.45&0.62\\
15510&0.711&-4.98&H96&&1.5&6.22&&5.21\\
15919&1.153&-4.6&M10&&0.53&1.14&&\\
16134&1.337&-4.61&M10&&3.85&1.21&&\\
16245&0.41&-4.92&S09&-5.96&&5.13&3.77&\\
16711&1.31&-4.573&G03&&&0.97&&\\
16852&0.575&-5.12&W04&&4.4&8.79&&3.50\\
17420&0.927&-5.04&M10&-5.05&0.6&7.35&0.99$^{a}$&11.35\\
17651&0.434&-5.039&S09&&16.25&7.33&&\\
18280&1.366&-4.539&G06&&&0.78&&\\
19849&0.82&-4.9&W04&-5.02&0.5&4.78$^{b}$&0.94&22.57\\
19884&1.115&-3.827&G03&&&0.001&&\\
19893&0.312&-4.545&G03&-5.54&&0.82&2.19&\\
21770&0.342&-4.569&G06&-5.08&47.8&0.95&1.05&\\
22263&0.632&-4.6&W04&-4.76&3.6&1.14$^{b}$&0.57&1.59\\
22449&0.484&-4.65&W04&-5.01&16.8&1.52&0.92&\\
23311&1.049&-5.29&M10&&1.7&10.91&&1.70\\
23452&1.43&-5.02&G03&&&6.98&&\\
23692&0.34&-4.64&G03&&14.8&1.44&&\\
25110&0.505&-4.957&G06&&&5.8&&\\
25544&0.755&-4.44&H96&&&0.4&&\\
25647&0.83&-4.372&G06&&&0.24$^{b}$&&\\
26394&0.6&-4.97&H96&&3.1&6.04&&3.21\\
27072&0.481&-4.77&H96&-5.71&9.84&2.8&2.77&\\
27887&0.935&-5.037&G06&&1.6&7.29&&1.91\\
29271&0.714&-4.94&H96&&1.7&5.49&&4.50\\
29568&0.713&-4.4&W04&-4.33&9.63&0.3&0.23&\\
31634&1.456&-4.696&G06&&&1.95&&\\
32439&0.525&-4.93&D91&&6.63&5.31&&1.65\\
32480&0.575&-4.85&W04&&4.8&3.95&&1.87\\
33277&0.573&-4.94&W04&&2.7&5.49&&4.34\\
34017&0.595&-4.96&W04&&2.8&5.85&&3.60\\
34065&0.624&-4.93&H96&&1.6&5.31&&8.93\\
34834&0.324&-4.905&G06&-5.88&50.7&4.86&3.46&\\
35136&0.576&-4.95&W04&&1.89&5.67&&7.86\\
35550&0.374&-4.08&D91&&129.7&0.02&&\\
36366&0.32&&&-5.41&49.9&&1.79&\\
36439&0.47&-5.31&D91&&5.88&11.03&&\\
37279&0.432&-4.72&D91&-6.12&5.7&2.2&4.51&\\
37288&1.379&-4.72&M10&&&2.2&&\\
37349&0.891&-4.39&W04&-4.9&2.5&0.28&0.74&0.95\\
38382&0.6&-5.012&G03&&&6.83&&\\
38784&0.719&-4.84&W04&&4.27&3.79&&0.69\\
40035&0.488&-4.81&M10&&11.26&3.34&&\\
40693&0.754&-4.95&W04&&0.3&5.67&&76.29\\
40702&0.413&-4.4&G03&-5.54&&0.3&2.19&\\
40843&0.487&-4.87&W04&&5.7&4.27&&\\
42438&0.618&-4.4&W04&&11.21&0.3$^{b}$&&0.21\\
43587&0.869&-5.04&W04&&2.5&7.35&&1.59\\
43726&0.661&-4.69&D91&-5.25&1.2&1.89$^{b}$&1.39&9.88\\
44075&0.521&-4.88&G03&&&4.44&&\\
44248&0.463&-4.55&G03&-5.26&21.17&0.84&1.42&\\
44722&1.418&-4.577&G06&&&1&&\\
44897&0.585&-4.62&D91&-4.76&3.9&1.28$^{b}$&0.57&1.91\\
45038&0.489&-4.845&G06&-6.1&7.55&3.87&4.44&\\
45170&0.731&-4.807&G06&&16.32&3.3&&\\
45333&0.605&-5.07&G03&&5.59&7.91&&0.89\\
45343&1.41&-4.521&G03&-4.67&&0.7&0.48&\\
46509&0.411&-4.64&G03&-4.97&&1.44&0.86&\\
46580&1.002&-4.5&M10&-4.66&3.1&0.61&0.47&0.50\\
46853&0.475&-4.956&G03&&8.77&5.78&&\\
47080&0.77&-4.73&D91&&6.97&2.31$^{b}$&&\\
47592&0.534&-4.86&M10&&5.7&4.11&&2.15\\
48113&0.619&-5.23&W04&&3&10.37&&4.95\\
49081&0.676&-5.06&W04&-4.83&3.4&7.72&0.65&1.60\\
49908&1.326&-5&M10&&4.21&6.6&&\\
50384&0.5&-4.792&G03&&2.16&3.09&&37.11\\
50505&0.653&-4.94&W04&&0.8&5.49&&20.12\\
50564&0.452&-4.749&G03&&&2.53&&\\
51459&0.541&-4.86&W04&-5.56&2.1&4.11&2.24&9.78\\
51502&0.399&-4.317&G06&-5&41.65&0.16&0.9&\\
53721&0.624&-5.02&W04&&2.8&6.98&&3.77\\
54646&1.255&-4.86&M10&&&4.11&&\\
54952&1.043&-4.657&G03&&&1.58&&\\
55203&0.606&-4.34&H96&&&0.19&&\\
55454&1.34&-4.317&G06&&&0.16$^{b}$&&\\
55846&0.778&-4.84&W04&&1.4&3.79&&4.92\\
56242&0.57&-4.94&W04&&3.3&5.49$^{b}$&&3.11\\
56809&0.566&-4.93&W04&&1.9&5.31&&8.50\\
56997&0.723&-4.55&W04&-5.17&2.4&0.84$^{b}$&1.22&1.93\\
56998&1.064&-4.527&G06&&&0.73&&\\
57443&0.664&-4.95&H96&&0.7&5.67&&24.32\\
57507&0.681&-4.92&H96&&&5.13&&\\
57757&0.518&-4.94&W04&-5.77&4&5.49&3.01&9.17\\
57939&0.754&-4.85&W04&&0.5&3.95$^{b}$&&17.86\\
58345&1.128&-4.452&G03&&&0.44&&\\
59199&0.334&&&-4.85&&&0.68&\\
59750&0.47&-4.44&D91&&&0.4&&\\
61053&0.568&-5.017&G03&&&6.92&&\\
61174&0.388&-4.959&G06&-5.62&60&5.83&2.47$^{a}$&\\
61317&0.588&-4.92&W04&&2.5&5.13&&4.12\\
61901&1.109&-5.003&G03&&3.6&6.66&&\\
61941&0.368&&&-5.13&31.36&&1.14&\\
62145&0.936&-4.72&S85&&&2.2&&\\
62207&0.557&-4.458&G03&&8.11&0.46&&0.53\\
62523&0.703&-4.58&W04&-5.1&2.8&1.01&1.09&1.73\\
62687&1.409&-4.516&G03&&&0.68&&\\
63366&0.769&-4.909&G06&&2.3&4.93&&\\
64241&0.455&-4.67&D91&-4.79&21.47&1.7&0.6&\\
64394&0.572&-4.76&W04&-5.69&4.4&2.67$^{b}$&2.69&1.79\\
64792&0.585&-4.4&W04&-4.45&7.4&0.3$^{b}$&0.3&0.88\\
64797&0.926&-4.76&D91&&2.86&2.67$^{b}$&&\\
64924&0.709&-5.04&W04&&2.2&7.35$^{b}$&&2.84\\
65352&0.78&-4.94&W04&&1.3&5.49&&3.95\\
66459&1.391&-5.076&G03&&&8.01&&\\
67090&1.42&-4.578&G06&&&1&&\\
67153&0.39&-4.887&G06&-5.02&63.9&4.55&0.93&\\
67275&0.508&-4.7&W04&-5.14&15&1.99$^{b}$&1.15&0.96\\
67308&1.322&-5.092&G06&&&8.3&&\\
67422&1.11&-4.68&M10&-4.92&0.3&1.79&0.77$^{a}$&42.42\\
67487&1.257&-4.727&G06&&&2.28&&\\
67620&0.703&-4.75&W04&&2.7&2.55&&1.99\\
67691&1.317&-4.987&G06&&&6.36&&\\
68184&1.04&-4.81&G06&&1.3&3.34&&3.04\\
68682&0.733&-5.04&H96&&2.2&7.35&&2.62\\
69671&0.596&-4.54&H96&&&0.79&&\\
69965&0.518&-4.641&G06&&&1.45&&\\
70218&1.275&-4.44&M10&&&0.4&&\\
70319&0.639&-4.94&W04&&1.1&5.49&&12.31\\
70497&0.497&-4.56&D91&-4.58&&0.9&0.39&\\
70857&0.774&-4.91&W04&&&4.95&&\\
71181&0.997&-5.37&D91&&1.8&11.19&&1.51\\
71284&0.364&-4.9&W04&-5.58&9.3&4.78&2.33&\\
71957&0.385&&&-5.12&&&1.11&\\
72567&0.576&-4.5&W04&-4.67&6.8&0.61&0.47&0.71\\
72659&0.72&-4.35&W04&-4.4&3.26&0.2$^{b}$&0.26&\\
72848&0.841&-4.52&M10&-4.78&4.5&0.69&0.59&0.45\\
73184&1.024&-4.63&M10&&2.6&1.36&&0.67\\
73695&0.647&-4.69&M10&-4.09&&1.89&0.13&\\
73996&0.429&-4.66&D91&-5.38&39.8&1.61&1.73&\\
74537&0.763&-5.131&G03&&2.12&8.97&&2.14\\
75277&0.804&-4.77&W04&&1.1&2.8&&5.85\\
75312&0.577&&&-5.67&&&2.64&\\
75718&0.788&-4.97&W04&&&6.04&&\\
76779&1.296&-4.78&M10&&3.21&2.93&&\\
77052&0.684&-4.83&W04&-5.35&11.01&3.64&1.63&0.21\\
77257&0.604&-4.97&W04&-6.21&3.1&6.04$^{b}$&4.94&3.98\\
77408&0.801&-4.39&W04&&&0.28$^{b}$&&\\
77760&0.563&-5.11&W04&&2.8&8.62&&9.90\\
77801&0.598&-4.85&W04&&1.2&3.95&&13.47\\
78072&0.478&-4.82&W04&&10.9&3.49&&\\
78459&0.612&-5.08&W04&-5.57&1.6&8.09&2.3&12.32\\
78775&0.734&-4.97&W04&&3.26&6.04&&\\
79607&0.599&-4.57&D91&-3.78&&0.95&0.06&\\
79755&1.409&-4.58&M10&-4.77&&1.01&0.58&\\
80686&0.555&-4.56&H96&-4.62&3.23&0.9&0.42&3.48\\
82003&1.31&-5.057&G03&&&7.67&&\\
82860&0.481&&&-5.08&&&1.05&\\
83389&0.728&-4.91&W04&&1.2&4.95&&6.79\\
83990&0.889&-4.703&G06&&0.2&2.02&&71.23\\
84862&0.619&-5.04&W04&&1.7&7.35&&8.48\\
85235&0.759&-4.93&W04&&6.79&5.31&&\\
85295&1.359&-4.72&M10&-4.93&3.21&2.2&0.79$^{a}$&\\
86036&0.602&-4.76&M10&-5.07&5.61&2.67&1.02$^{a}$&0.98\\
86201&0.43&-4.439&G06&-4.99&&0.4&0.89&\\
86486&0.415&-4.665&G06&-5.62&&1.65&2.44&\\
86614&0.434&-4.552&G03&-5.7&&0.85&2.74&\\
88745&0.528&-5.02&W04&&&6.98&&\\
88972&0.876&-4.97&W04&&2.1&6.04$^{b}$&&1.56\\
89042&0.592&-4.91&H96&&&4.95&&\\
89211&1.297&-4.97&W04&&&6.04&&\\
89348&0.44&&&-5.2&&&1.27&\\
89805&0.584&-4.61&H96&-4.76&&1.21&0.57&\\
89937&0.489&&&-6.22&6.01&&4.99&\\
93017&0.594&-4.88&D91&-5.75&6.8&4.44&2.94&0.82\\
94346&0.804&-4.52&W04&&&0.69$^{b}$&&\\
95149&0.628&-4.31&H96&&&0.15&&\\
96441&0.395&-4.51&G03&-5.31&6.5&0.65&1.54&\\
96895&0.643&-5.1&W04&&2.8&8.44&&3.66\\
98470&0.498&-4.472&G03&-3.99&72.6&0.5&0.1$^{a}$&0.10\\
98698&1.128&-4.496&G06&-5.1&&0.59$^{b}$&1.08&\\
98959&0.648&-4.86&H96&&1.2&4.11&&11.26\\
99137&0.53&-4.38&H96&-4.69&&0.26&0.49&\\
99461&0.868&-4.98&H96&&2.7&6.22&&0.91\\
99701&1.431&-4.551&G03&-4.98&&0.85&0.86&\\
99825&0.878&-4.555&G03&&2&0.87&&1.77\\
100925&0.724&-4.9&H96&&0.4&4.78&&53.79\\
101997&0.719&-4.92&W04&&1.2&5.13&&6.40\\
102040&0.611&-4.92&W04&&&5.13&&\\
102186&1.324&-4.467&G06&&&0.49&&\\
102485&0.426&-4.832&G03&-5.07&34.94&3.67&1.03&\\
103389&0.507&-4.402&G06&-4.67&&0.3&0.47&\\
104092&1.119&-4.597&G06&&14.33&1.12&&\\
104239&0.901&-4.74&D91&-4.52&2.1&2.43&0.35&1.41\\
104440&0.59&-5.099&G03&-4.93&&8.43&0.8$^{a}$&\\
105090&1.397&-4.751&G06&-5.29&&2.56&1.5&\\
105312&0.737&-4.87&H96&&&4.27&&\\
105712&0.723&-4.39&H96&&&0.28&&\\
105858&0.494&-4.998&G06&&9.84&6.56&&\\
106696&0.879&-4.499&G06&-5.23&2&0.6&1.35&1.43\\
107310&0.512&-4.94&D91&-5.58&5.45&5.49&2.31&8.96\\
107350&0.587&-4.44&D91&-4.45&10.6&0.4$^{b}$&0.3&0.31\\
107556&0.18&-4.768&G06&-5.25&&2.77&1.39&\\
107649&0.601&-4.8&H96&-5.62&2.4&3.2&2.46&4.29\\
108870&1.056&-4.56&H96&&&0.9&&\\
109176&0.435&-4.357&G06&-5.8&5.42&0.22&3.11&\\
109422&0.489&-4.99&W04&&13.7&6.41&&\\
110109&0.614&-4.86&H96&&0.32&4.11&&133.90\\
110443&1.329&-4.839&G06&&&3.78&&\\
110649&0.665&-5.07&H96&&2.8&7.91&&5.52\\
111449&0.446&-4.606&G06&-5.07&33.31&1.18&1.02&\\
111960&1.143&-4.851&G06&&&3.96&&\\
112117&0.584&-4.89&H96&&4.5&4.61&&1.59\\
112447&0.501&-5.07&W04&&8.9&7.91&&6.00\\
112774&1.452&-4.486&G06&&&0.55&&\\
113283&1.094&-4.26&H96&-4.56&&0.1$^{b}$&0.38&\\
113357&0.666&-5.08&W04&&2.6&8.09$^{b}$&&3.23\\
113421&0.744&-5.08&W04&&&8.09&&\\
113576&1.379&-4.907&G06&&3.21&4.9&&\\
114361&1.201&-4.617&G03&&&1.26&&\\
114948&0.521&-4.94&G06&-4.46&&5.49&0.3$^{a}$&\\
114996&0.41&-4.67&G06&-6.12&&1.7&4.52&\\
116215&1.29&-5.079&G06&&&8.07&&\\
116745&0.989&-5&G03&&2.4&6.6&&0.86\\
116763&0.802&-4.73&H96&&1.3&2.31&&4.30\\
116771&0.507&-4.98&D91&-6.19&6.8&6.22&4.84&5.05\\
562452&0.811&-4.86&H96&&0.7&4.11&&10.81\\
GJ 750A&1.706&-4.628&G06&&&1.34&&\\
HD 10361&0.984&-4.9&H96&&&4.78&&\\
HD 202275&0.529&-4.905&G03&&&4.86&&\\
HD 212698&0.677&-4.853&G06&&&4&&
\enddata
\tablecomments{$^{a}$=There is a large discrepancy between X-ray age and chromospheric age; these stars are discussed in detail in the text. $^{b}$ = Gyrochronology age (preferred to chromospheric and X-ray age) is listed in Table 1.\\
Reference notes: D91 = Duncan (1991), M10=Maldonado et al. (2010), G06 = Gray et al. (2006), G03 = Gray et al. (2003), W04 = Wright et al. (2004), H96 = Henry et al. (1996), S85 = Soderblom et al. (1985), S09 = Schroeder et al. (2009)}
\end{deluxetable}

\clearpage

\begin{deluxetable}{ccccc}
\tablewidth{0pt}
\tablecaption{DEBRIS A Star Ages}
\tablehead{
\colhead{HIP}&
\colhead{Isochrone Age}&
\colhead{Isochrone Source}\\
\colhead{}&
\colhead{(Gyr)}&
\colhead{}}
\startdata
1473&0.45&YREC\\
2072&0.72&YREC\\
4436&0.6&R05\\
8903&0.55&YREC\\
10064&0.73&Li-Han\\
10670&0.16&YREC\\
11001&0.405&R05\\
12225&0.55&YREC\\
12413&0.37&YREC\\
12706&0.5&YREC\\
12832&0.58&YREC\\
14146&0.7&YREC\\
14293&0.07&YREC\\
15197&0.8&YREC\\
17395&0.39&Li-Han\\
23875&0.39&Li-Han\\
27288&0.23&YREC\\
32607&0.7&Li-Han\\
35350&0.55&YREC\\
36850&0.25&YREC\\
41307&0.17&YREC\\
44127&0.75&YREC\\
44382&0.42&YREC\\
44901&0.8&YREC\\
45493&0.71&YREC\\
45688&0.33&YREC\\
48390&0.45&YREC\\
49593&0.75&YREC\\
50191&0.41&YREC\\
50372&0.38&YREC\\
50888&0.07&YREC\\
51658&0.42&YREC\\
53910&0.31&YREC\\
53954&0.32&YREC\\
54872&0.69&Li-Han\\
55705&0.57&YREC\\
57328&0.48&YREC\\
57632&0.1&YREC\\
58001&0.4&Li-Han\\
58684&0.15&Li-Han\\
59774&0.49&Li-Han\\
60965&0.26&YREC\\
61468&0.81&YREC\\
61622&0.31&YREC\\
61932&0.45&Li-Han\\
61960&0.5&YREC\\
63125&0.19&YREC\\
65109&0.26&YREC\\
65378&0.37&YREC\\
66249&0.49&YREC\\
69713&0.04&YREC\\
69732&0.29&YREC\\
72220&0.29&YREC\\
72622&0.65&YREC\\
75695&0.81&YREC\\
75761&0.07&YREC\\
76267&0.27&YREC\\
76952&0.4&YREC\\
77622&0.52&Li-Han\\
83613&0.4&YREC\\
84379&0.35&Li-Han\\
85829&0.7&YREC\\
88771&0.55&YREC\\
88866&0.8&YREC\\
92024&0.01&Li-Han\\
93408&0.59&YREC\\
93506&0.48&YREC\\
94083&0.1&Li-Han\\
95853&0.45&YREC\\
97534&0.58&YREC\\
97649&0.7&YREC\\
98495&0.25&YREC\\
102333&0.52&YREC\\
105319&0.6&YREC\\
109285&0.39&Li-Han\\
109427&0.5&YREC\\
110935&0.06&YREC\\
111188&0.18&YREC\\
112623&0.6&Li-Han\\
116758&0.6&YREC\\
116928&0.7&YREC\\
117452&0.22&YREC

\enddata
\tablecomments{R05=Rieke et al. (2005)\\DEBRIS star HIP 88726 (not included in this table) is a member of the beta Pic moving group and thus is $\sim$12 Myr old (Zuckerman $\&$ Song (2004)).}
\end{deluxetable}

\begin{figure}
\begin{center}
\title{Gyrochronology Ages of DEBRIS FGK Stars}
\plotone{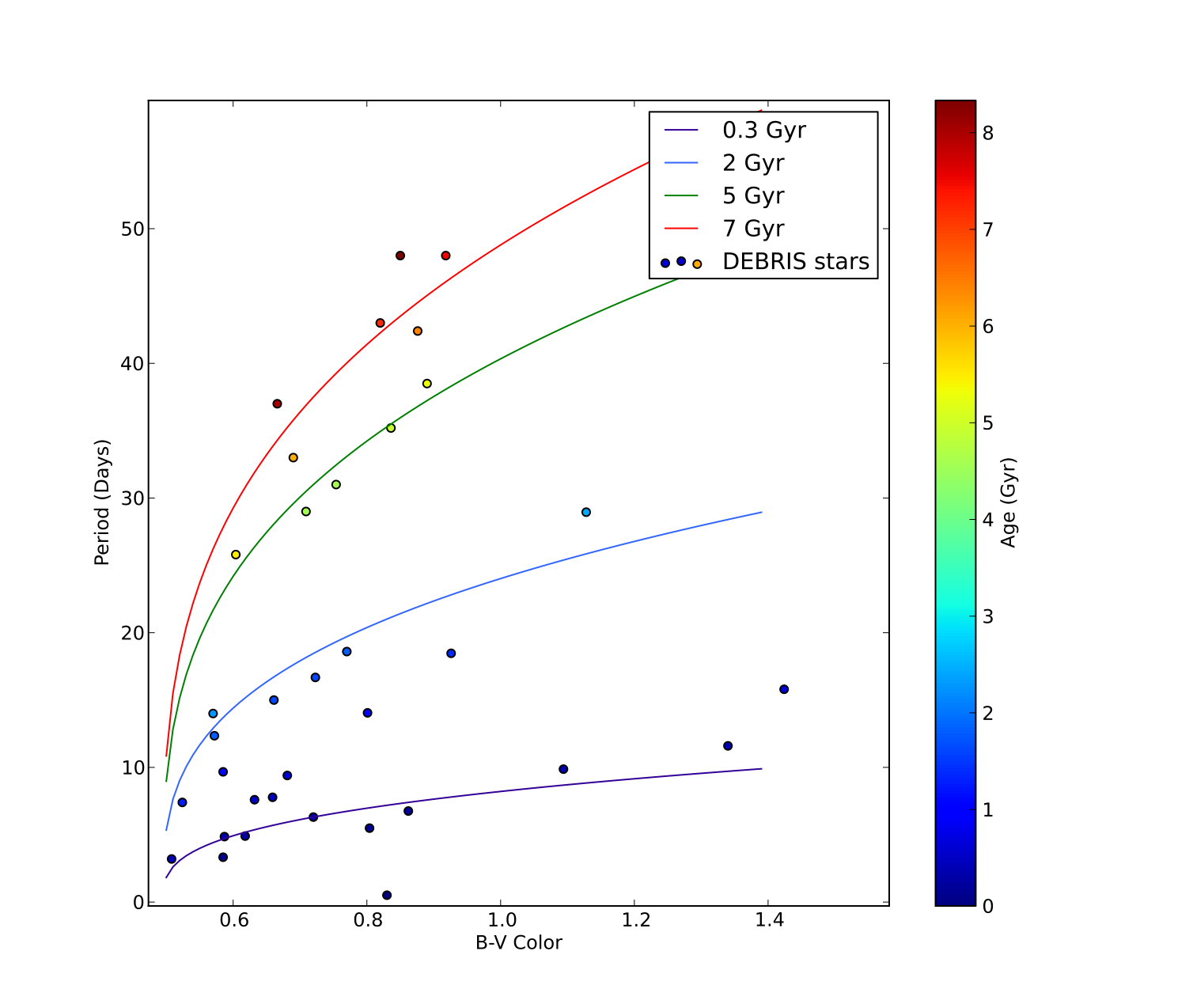}
\caption{Ages of DEBRIS stars using gyrochronology, plotted in color-rotation space. Overplotted are four ``gyrochrones" using Equation 1.\label{fig1}}
\end{center}
\end{figure}

\begin{figure}
\begin{center}
\title{Comparison of P$_{rot}$ and vsin{i}}
\plotone{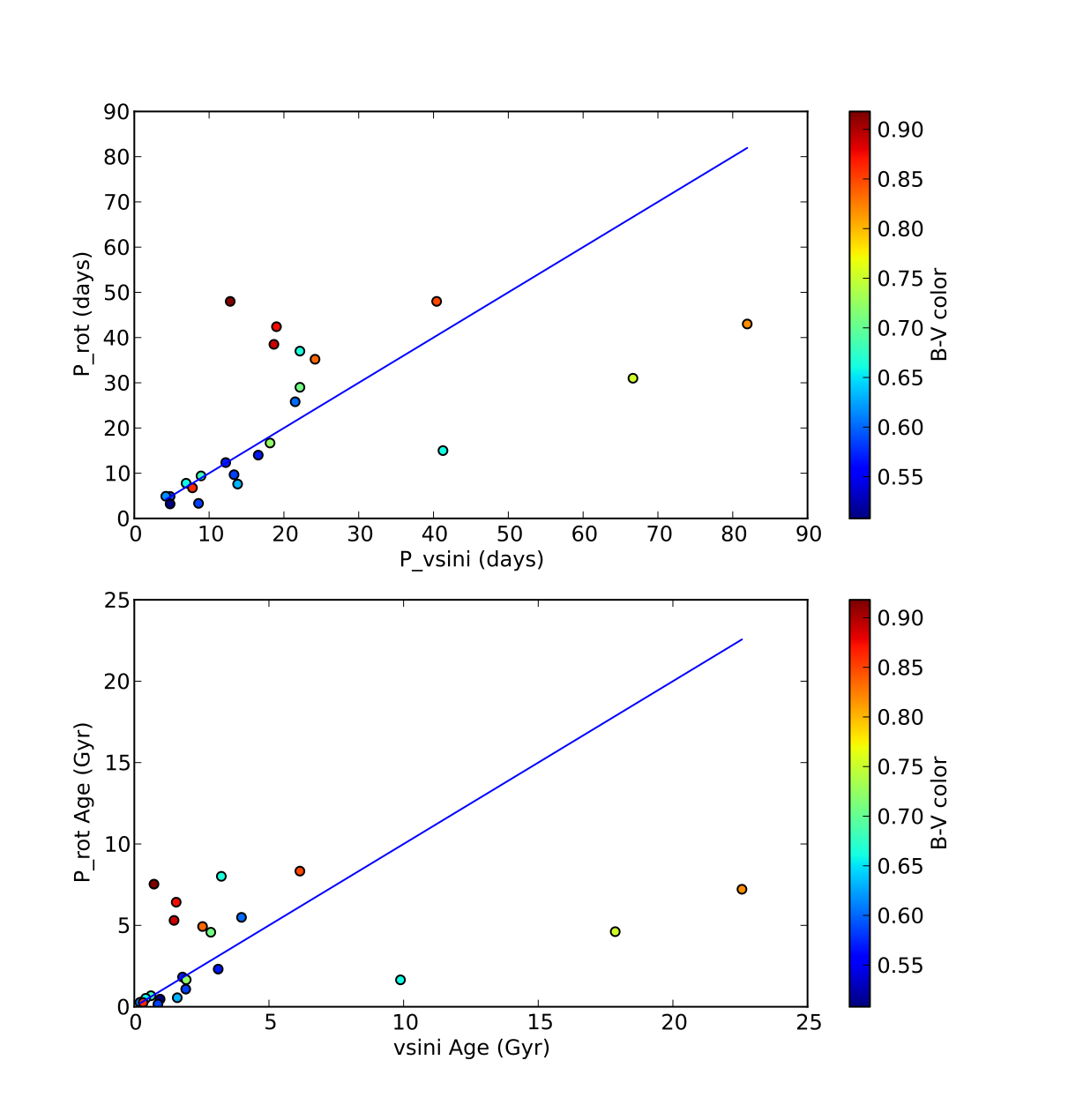}
\caption{$Upper Panel:$ Direct comparison of measured (empirical) P$_{rot}$ to the rotational period calculated from vsin$i$. The blue line represents the 1:1 relation. It seems that there is more scatter in the relation for late-type stars, and for stars with longer rotation periods. $Lower Panel:$ Comparison between gyrochronology ages calculated using vsin$i$ and P$_{rot}$. Again, we find significant scatter for late-type stars, and at long P$_{rot}$.\label{fig2}}
\end{center}
\end{figure}

\begin{figure}
\center
\plotone{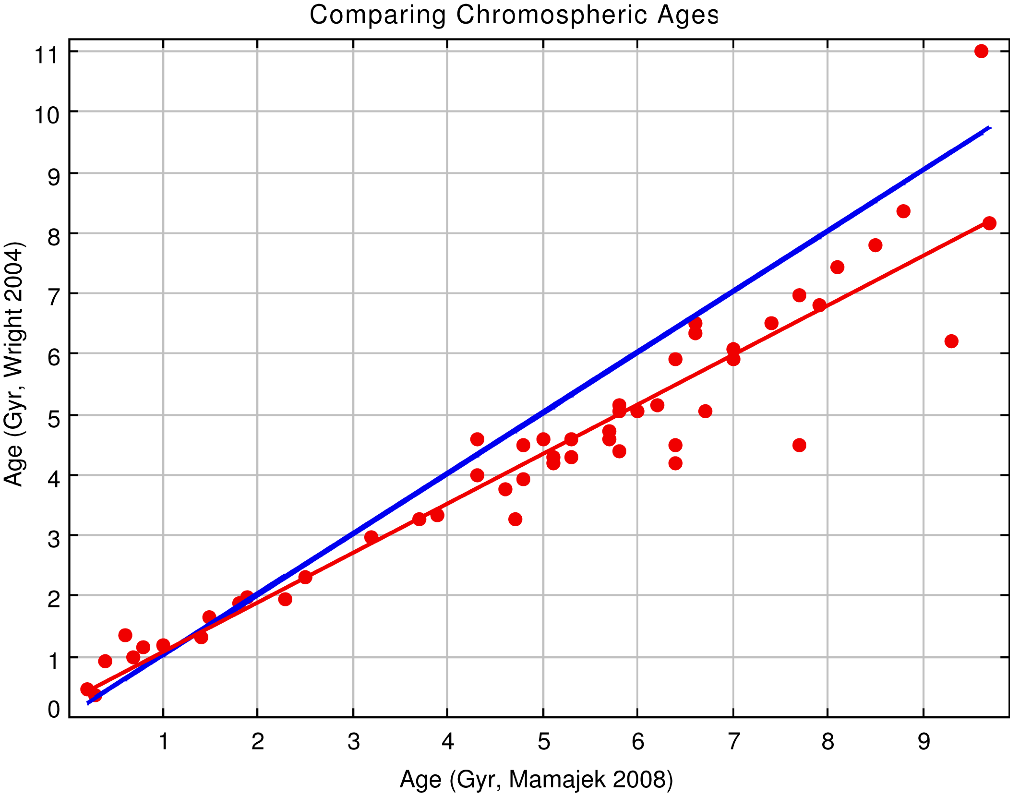}
\caption{Comparison between ages calculated by Wright et al. (2004) and Mamajek $\&$ Hillenbrand (2008). The blue line represents the 1:1 relation. MH08 tends to calculate older ages, and we follow their equation for stellar age.\label{fig3}}
\end{figure}

\begin{figure}
\center
\title{Chromospheric Age vs. Gyrochronology Age}
\plotone{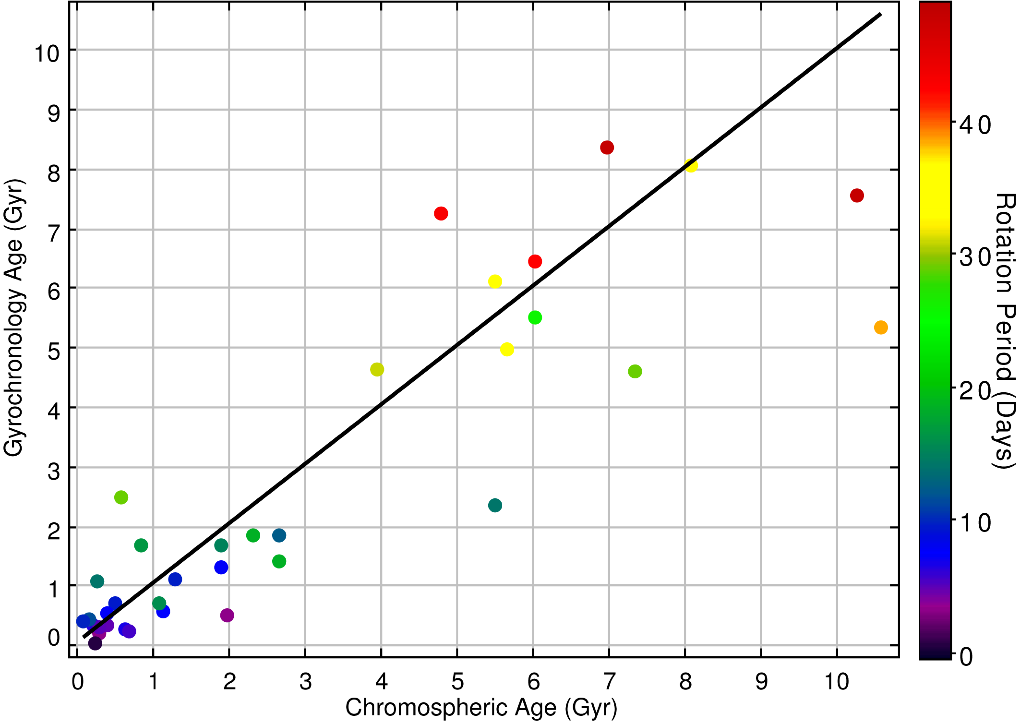}
\caption{Our results show a fairly good correlation between ages determined by R'HK and by rotational period (our analysis suggests an average discrepancy of 87$\%$. Period-based ages are less reliable for slow rotators, because the differential spin-down is not as apparent. \label{fig4}}
\end{figure}

\begin{figure}
\center
\title{X-ray Age vs. Gyrochronology Age}
\plotone{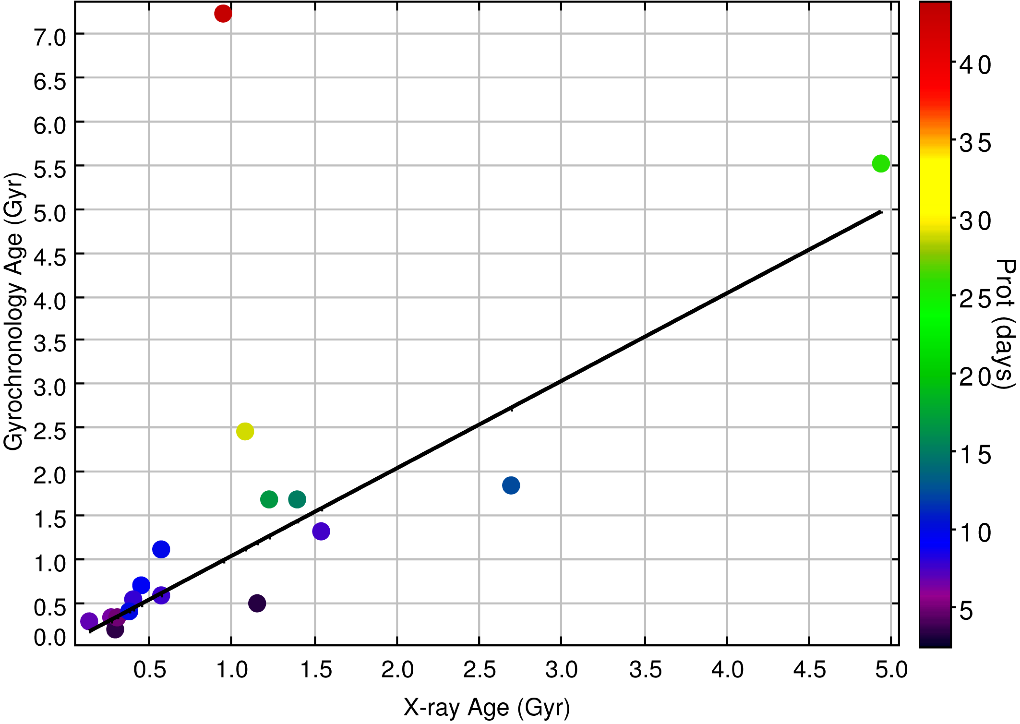}
\caption{From our data, it seems that X-ray ages are in good agreement with gyrochronology ages. The outlier at high gyrochonology age (HIP 19849) is a very slow rotator, thus its gyrochronology age should not be trusted. \label{fig5}}
\end{figure}

\begin{figure}
\center
\title{X-ray Age vs. Chromospheric Age}
\plotone{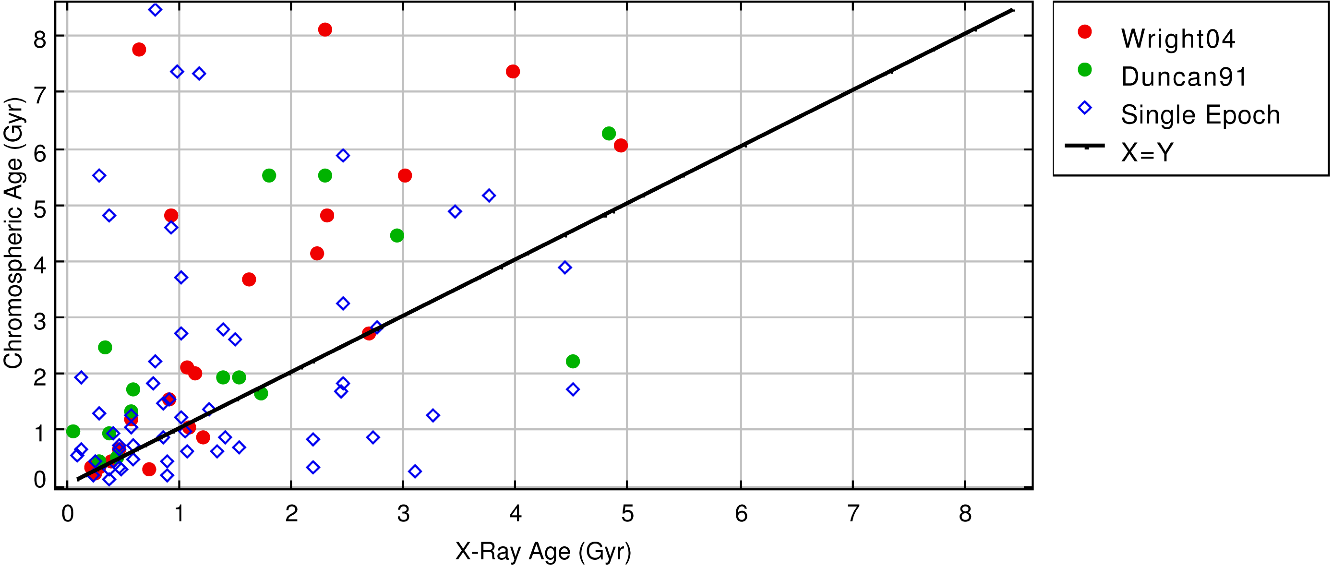}
\caption{This figure compares ages derived using chromospheric activity ($R^\prime_{HK}$), and ages derived using X-ray data from the ROSAT all-sky survey. We separated the data based on the type of survey the $R^\prime_{HK}$ value came from. Data from single epoch surveys yield less reliable chromospheric ages. These are the blue diamonds in our figure.  Data from two large multi-epoch surveys (Wright et al. (2004) and Duncan (1991)) are represented here by red and green data points, respectively. It is clear from this plot that chromospheric ages are, in general, older than X-ray ages. Also, as expected, many of the outlying data points are from single epoch chromospheric emission surveys. \label{fig6}}
\end{figure}

\begin{figure}
\center
\title{Histogram of A star Ages}
\plotone{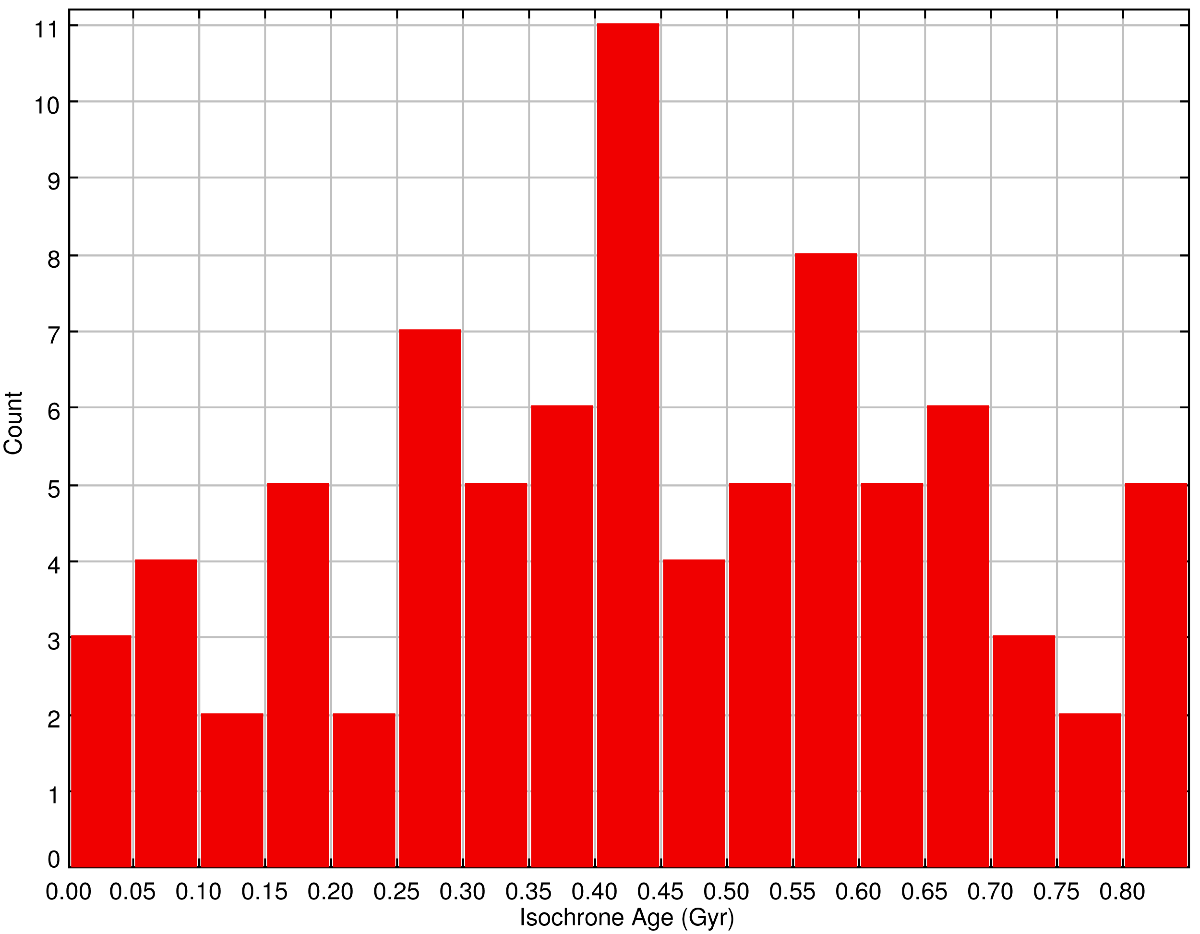}
\caption{Histogram of 83 DEBRIS A star ages. We used the YREC isochrones when possible, and when the log(g) and Teff values were out of the range of the YREC isochrones, we used isochrones from Li $\&$ Han (2008). When neither of our isochrones were appropriate, we used isochrone ages from Rieke et al. (2005). All of the A stars are younger than 1 Gyr.\label{fig7}}
\end{figure}


\begin{thebibliography}{}

\bibitem[Allende Prieto (1999)]{All99}Allende Prieto, C. $\&$ Lambert, D. L., 1999, \aap, 352, 555.
\bibitem[Anderson (2011)]{And11}Anderson, E., $\&$ Francis, C., 2011, arXiv:1108.4971v1.
\bibitem[Baliunas (1983)]{Bal83}Baliunas, S. L., Hartmann, L., Noyes, R. W., et al., 1983, \apj, 275, 752.
\bibitem[Baliunas (1996)]{Bal96}Baliunas, S., Sokoloff, D., Soon, W., 1996, \apj, 457, 99.
\bibitem[Barnes (2003)]{Bar03}Barnes, S. A., 2003, \apj, 586, 464.
\bibitem[Barnes (2007)]{Bar07}Barnes, S. A., 2007, \apj, 154, 153.
\bibitem[Barnes (2010)]{Bar10}Barnes, S. A., 2010, \apj, 722, 222.
\bibitem[Baumann (2010)]{Bau10}Baumann, P., Ram$\acute{i}$rez, I., Mel$\acute{e}$ndez, J., Asplund, M., Lind, K., 2010, \aap, 519A, 87.
\bibitem[Bertelli (2009)]{Ber09}Bertelli, G., Nasi, E., Girardi, L., Marigo, P., 2009, \aap, 508, 335.
\bibitem[Booth (2009)]{Boo09}Booth, M., Wyatt, M. C., Morbidelli, A., Moro-Mart$\acute{i}$n, A., Levison, H. F., 2009, \mnras, 399, 385.
\bibitem[Chen (2001)]{Che01}Chen, Y. Q., Nissen, P. E., Benoni, T., Zhao, G., 2001, \aap, 371, 943.
\bibitem[Donahue (1993)]{Don93}Donahue, R. A., 1993, \pasp, 105, 804.
\bibitem[Donahue (1996)]{Don96}Donahue, R. A., Saar, S. H., Baliunas, S. L., 1996, \apj, 466, 384.
\bibitem[Duncan (1991)]{Dun91}Duncan, D. K., Vaughan, A. H., Wilson, O. C., et al., 1991, \apjs, 76, 383.
\bibitem[Gerbaldi (1999)]{Ger99}Gerbaldi, M., Faraggiana, R., Burnage, R., et al., 1999, \aaps, 137, 273.
 \bibitem[Gray (2003)]{Gra03}Gray, R. O., Corbally, C. J., Garrison, R. F., McFadden, M. T., Robinson, P. E., 2003, \aj, 126, 2048.
 \bibitem[Gray (2006)]{Gra06}Gray, R. O., Corbally, C. J., Garrison, R. F., et al., 2006, \aj, 132, 161.
\bibitem[Heng (2010)]{Hen10}Heng \& Tremaine, 2010, \mnras, 401, 867.
\bibitem[Henry (1996)]{Hen96}Henry, T. J., Soderblom, D. R., Donahue, R. A., Baliunas, S. L., 1996, \aj, 111, 439.
\bibitem[King (2003)]{Kin03}King, J. R., \& Schuler, S. C., 2003, \aj, 125, 1980.
\bibitem[Lafrasse (2010)]{Laf10}Lafrasse, S., Mella, G., Bonneau, D., et al., 2010, SPIE Conference Series, 7734, 140.
\bibitem[Lepine (2007)]{Lep07}Lepine, S., \& Bongiorno B., 2007, \aj, 133, 889.
\bibitem[Li (2008)]{Li08}Li, Z., \& Han, Z., 2008, \mnras, 387, 105.
\bibitem[Maldonado (2010)]{Mal10}Maldonado, J., Martinez-Arnaiz, R. M., Eiroa, C., Montes, D., Montesinos, B., 2010, arXiv:1011.3983.
\bibitem[Mamajek \& Hillenbrand (2008)]{MaMH08}Mamajek, E. E., \& Hillenbrand, L. A., 2008, \apj, 687, 1264.
\bibitem[Matthews (2010)]{Mat10}Matthews, B. C., Sibthorpe, B., Kennedy, G., et al., \aap, 518, L135.
\bibitem[Noyes (1984)]{Noy84}Noyes, R. W., Hartmann, L. W., Baliunas, S. L., Duncan, D. K., Vaughan, A. H., 1984, \apj, 279, 763.
\bibitem[Perryman (1997)]{Per97}Perryman, M. A. C., Lindegren, L., Kovalevsky, J., et al., 1997, \aap, 323, 49.
\bibitem[Phillips (2010)]{Phi10}Phillips, N. M., Greaves, J. S., Dent, W. R. F., et al., 2010, \mnras, 403, 1089.
\bibitem[Pilbratt (2010)]{Pil10}Pilbratt, G. L., Riedinger, J. R., Passvogel, T., et al. 2010, \aap, 518L, 1P.
\bibitem[Pinsonneault (2004)]{Pin04}Pinsonneault, M. H., Terndrup, D. M., Hanson, R. B., Stauffer, J. R., 2004, \apj, 600, 946.
\bibitem[Raghavan (2010)]{Rag10}Raghavan, D., McAlister, H. A., Henry, T. J., et al., 2010, \apjs, 190, 1.
\bibitem[Rieke (2005)]{Rie05}Rieke, G. H., Su, K. Y. L., Stansberry, J. A., et al., 2005, \apj, 620,1010.
\bibitem[Rutten (1987)]{Rut87}Rutten, R. G. M., 1987, \aap, 177, 131.
\bibitem[Sibthorpe (in prep)]{SibIP}Sibthorpe et al. 2012, in prep.
\bibitem[Soderblom (1985)]{Sod85}Soderblom, D. R., 1985, \aj, 90, 2103.
\bibitem[Song (2001)]{Son01}Song, I., et al., Caillault, J. P., Barrado y Navascue$\acute{e}$s, D., Stauffer, J. R., 2001, \apj, 546, 352.
\bibitem[Song (2004)]{Son04}Song, I., Zuckerman, B., Bessell, M. S., 2004, \apj, 614, 125.
\bibitem[Soubiran (2010)]{Sou10}Soubiran, C., Le Campion, J. F., Cayrel de Strobel, G., Caillo, A., 2010, \aap, 515A, 111.
\bibitem[Su (2006)]{Su06}Su, K. Y. L., Rieke, G. H., Stansberry, J. A., et al. 2006, \apj, 653, 675.
\bibitem[Takeda (2007)]{Tak07}Takeda, Y., 2007, \pasj, 59, 335.
\bibitem[Thureau (in prep)]{ThuIP}Thureau et al. 2012, in prep.
\bibitem[Turon (1993)]{Tur93}Turon, C., Creze, M., Egret, D., et al., 1993, BICDS, 43, 5.
\bibitem[Valenti (2005)]{Val05}Valenti, J. A., \& Fischer, D. A., 2005, \apjs, 159, 141.
\bibitem[Voges (1999)]{Vog99}Voges, W., Aschenbach, B., Boller, Th., et al., 1999, \aap, 349, 389.
\bibitem[Wright (2004)]{Wri04}Wright, J. T., Marcy, G. W., Butler, R. P., Vogt, S. S., 2004, \apjs, 152, 261.
\bibitem[Wright (2011)]{WriIP}Wright, N. J., Draks, J. J., Mamajek, E. E., Henry, G. W., 2011, \apj, 743, 48.
\bibitem[Wyatt (2008)]{Wya08}Wyatt, M. C., 2008, \araa, 46, 339.
\bibitem[Yi (2003)]{Yis03}Yi, S. K., Kim, Y.-C., Demarque, P., 2003, \apjs, 144, 259.
\bibitem[Zuckerman (2001)]{Zuc01}Zuckerman, B., 2001, \araa, 39, 549.
\bibitem[Zuckerman (2004)]{Zuc04}Zuckerman, B. \& Song, I., 2004, \araa, 42, 685.
\end{thebibliography}
\end{document}